# Effects of Graphene/BN Encapsulation, Surface Functionalization and Molecular Adsorption on the Electronic Properties of Layered InSe: A First-Principles Study

Andrey A. Kistanov,[a, b] Yongqing. Cai,*[, b] Kun Zhou,*[, a] Sergey V. Dmitriev[c, d] and Yong-Wei Zhang*[b]

## Abstract

By using first-principles calculations, we investigated the effects of graphene/boron nitride (BN) encapsulation, surface functionalization by metallic elements (K, Al, Mg and typical transition metals) and molecules (tetracyanoquinodimethane (TCNQ) and tetracyanoethylene (TCNE)) on the electronic properties of layered indium selenide (InSe). It was found that an opposite trend of charge transfer is possible for graphene (donor) and BN (acceptor), which is dramatically different from phosphorene where both graphene and BN play the same role (donor). For InSe/BN heterostructure, a change of the interlayer distance due to an out-of-plane compression can effectively modulate the band gap. Strong acceptor abilities to InSe were found for the TCNE and TCNQ molecules. For K, Al and Mg-doped monolayer InSe, the charge transfer from K and Al atoms to the InSe surface was observed, causing an n-type conduction of InSe, while *p*-type conduction of InSe observed in case of the Mg-doping. The atomically thin structure of InSe enables the possible observation and utilization of the dopant-induced vertical electric field across the interface. A proper adoption of the *n*- or *p*-type dopants allows for the modulation of the work function, the Fermi level pinning, the band bending, and the photo-adsorbing efficiency near the InSe surface/interface. Investigation on the adsorption of transition metal atoms on InSe showed that Ti-, V-, Cr-, Mn-, Co-adsorbed InSe are spin-polarized, while Ni-, Cu-, Pd-, Ag- and Au-adsorbed InSe are non-spin-polarized. Our results shed lights on the possible ways to protect InSe structure and modulate its electronic properties for nanoelectronics and electrochemical device applications.

## Introduction

Indium selenide (InSe), a newly emerging two-dimensional (2D) layered metal monochalcogenide III-VI compound, has attracted great attention owing to its high carrier mobility (up to $10^4$ cm$^2$ V$^{-1}$s$^{-1}$ at liquid-helium temperature) and anomalous optical response.[1,2] In contrast to transition metal dichalcogenides (TMDs) such as MoS$_2$, which possess a direct band gap for monolayer and an indirect band gap for multilayer,[3, 4] InSe has a direct band gap for bulk phase and an indirect band gap obtained by reducing the layer number to make its thickness below a critical value at about several nanometers.[5, 6] In addition, InSe samples with a direct band gap span a broad thickness range (from infinite thickness down to several nanometers), allowing a stronger excitonic emission and a broader frequency spectrum than TMDs. As a result, layer engineering of InSe should be highly effective for tuning the momentum conservation of quasi-particles, in addition to the traditional quantum confinement effect.

Recently, much effort has been devoted to investigating the exfoliation[5, 6] and electronic applications of InSe, for example, in optoelectronics[8-12] and photovoltaics.[13-17] However, the mobility of InSe-based transistors at ambient condition was found to degrade due to environmental effects and defects.[18, 19] This high chemical sensitivity of InSe can be attributed to the lone-pair states of surface Se atoms.[20,21] Therefore, identifying effective strategies to protect InSe sheets is important for the reliability of InSe-based electronic devices.[22, 23] For atomically thin 2D materials like $MoS_2$ and phosphorene, encapsulation by chemically more inert 2D layers, like graphene and BN, was found to effectively suppress the adverse environment effect[24-27] and rectify interfacial resistance at the electrode contact.[28] Concerning InSe, the effects of BN and graphene encapsulation on its electronic and chemical properties remain unclear.

For practical applications of a semiconducting InSe layer, there is a need to develop methods to control the polarity and concentration of its conducting carriers. As an intrinsic InSe layer shows a *p*-type conduction,[20] the realization of *n*-type conduction is highly desired. Traditional substitutional doping strategy that has been widely adopted in bulk semiconductors is highly challenging for 2D materials as it tends to destroy their 2D structural integrity. To this end, various doping approaches by using surface functionalization of 2D materials have been developed.[29] Functionalization of graphene/h-BN heterostructure by hydrogen or fluorine atoms leads to an increase of the band gap, giving rise to an n- or p-type doped graphene,[30] respectively. Furthermore, typical donor (*n*-type) and acceptor (*p*-type) organic molecules, such as tetracyanoquinodimethane (TCNQ) and tetracyanoethylene (TCNE), can dramatically modify the band gap of graphene and other common 2D materials.[31-33] Recently, it was reported that *in-situ* deposition of potassium (K) atoms is able to significantly tune the band gap of black phosphorus[34] and intercalated K dopants are able to increase the electron mobility due to a charge-transfer induced giant vertical electrical field.[35] However, to the best of our knowledge, no study has been conducted on the modulation of electronic properties of InSe through surface functionalization and molecular adsorption.

In this work, by using *ab initio* first-principles calculations, we systematically investigated the electronic structure of monolayer InSe through graphene/BN encapsulation, chemical functionalization and molecular adsorption. We found that an opposite trend of charge transfer is possible for graphene (donor) and BN (acceptor). Also, there is a strong dependence between the interlayer distance of InSe/BN heterostructure and the band gap. We explored the charge-transfer induced effect caused by the adsorption of K, Al and Mg atoms on monolayer InSe, InSe/graphene and InSe/BN heterostructures. For all the three cases, the charge transfer from the K, Al and Mg atoms to the InSe surface nearly linearly decreases with an increasing distance from the adsorbed atom to the surface.

We also studied the effects of adsorption of transition metals (Ti, V, Cr, Mn, Co, Ni, Cu, Pd, Ag and Au) on the electronic properties of InSe. It was found that the adsorption of these considered transition metal atoms results in significant lattice distortions. Moreover, Ti-, V-, Cr-, Mn- and Co-adsorbed InSe are spin-polarized, while Ni-, Cu-, Pd-, Ag- and Au-adsorbed InSe are non-spin-polarized. In addition, TCNE and TCNQ molecules adsorbed on monolayer InSe show a strong acceptor ability which decreases with an increasing adsorption distance between molecules and the surface. The findings revealed here provide valuable insights into the modulation of electronic properties of monolayer InSe, which may be important for the fabrication of InSe-based nanodevices.

**Computational details**

Spin-polarized density functional theory (DFT) calculations were performed by using VASP[36] package. Perdew−Burke−Ernzerhof (PBE)[37] exchange-correlation functional under the generalized gradient approximation (GGA) was selected. Corrected functional with Becke88 optimization (optB88) was used for considering the van der Waals (vdW) interaction and the dipole correction was considered in our calculations. The relaxed lattice constant of monolayer InSe was $a = b = 4.102$ Å. To simulate the surface chemical functionalization, a $3 \times 3 \times 1$ supercell of InSe was used. A vacuum space of 25 Å was introduced along the out-of-plane direction. The first Brillouin zone was sampled with a $6 \times 6 \times 1$ k-mesh grid. The kinetic energy cutoff of 400 eV was adopted. All structures were fully relaxed until the total energy and atomic forces were smaller than $10^{-5}$ eV and 0.01 eV/Å, respectively. The absorption energy ($E_a$) of a molecule on the monolayer InSe surface was calculated as $E_a = E_{Mol+InSe} - E_{InSe} - E_{Mol}$, where $E_{Mol+InSe}$, $E_{InSe}$ and $E_{Mol}$ are the energies of the molecule-adsorbed InSe, monolayer InSe surface, and molecule, respectively. The binding strength $\sigma_b$ was calculated as $\sigma_b = E_b/S_{surface}$, where $E_b$ is the binding energy for the InSe/graphene or the InSe/BN heterostructures and $S_{surface}$ is the surface area of the supercell. The electronic charge analysis was performed by the Bader approach.[38]

**Results and discussion**

**Heterostructures formed by graphene and BN encapsulation.** To suppress the potential environmental effect, passivating the InSe with the relatively inert graphene or BN layer could be an effective approach, which has been demonstrated in phosphorene flakes. However, the effect of graphene and BN on the electronic properties of InSe remains still unclear. Herein, commensurate atomic structures of the vertical InSe/graphene and InSe/BN heterostructures through stacking a monolayer InSe and a monolayer graphene/BN along the normal direction were created by using the $3 \times 3 \times 1$ supercell of InSe and the $5 \times 5 \times 1$ supercell of graphene/BN. The in-plane lattice constant of the hybrid structures was adjusted to the lattice constant of the InSe supercell, and in the mismatch strain in graphene/BN layer was smaller than 1.6 %. The optimized distances of InSe from graphene and BN were 3.37 and 3.48 Å for InSe/graphene and InSe/BN, respectively. Figures 1a–c (upper panel) show the atomic structures of the optimized monolayer InSe, InSe/graphene and InSe/BN heterostructures. The binding strengths $\sigma_b$ for InSe/graphene and InSe/BN heterostructures were 0.52 and 0.23 eV/Å$^2$, respectively.

Figure 1 (lower panel) presents the band structures of monolayer InSe, InSe/graphene and InSe/BN heterostructures. It is found that monolayer InSe is a semiconductor (Figure 1a, lower panel) with an indirect band gap of 1.33 eV, which is consistent with previous works.[20, 39] For InSe/graphene heterostructure (Figure 1b), the Dirac cone formed by graphene is preserved and almost overlaps the conduction band edges of InSe. Instead, InSe/BN heterostructure (Figure 1c, lower panel) remains a semiconductor with the reduced indirect band gap of 1.26 eV, compared with the naked monolayer InSe. The top valence band of BN is aligned with that of InSe, while there are no BN states overlapping with the InSe state in the lower lying conduction bands due to the much larger band gap of BN.

In order to gain further insight into the interlayer interactions of the considered heterostructures, we examined the charge transfer between InSe and graphene/BN layers. The differential charge density (DCD) $\Delta\rho(r)$ is calculated as $\Delta\rho(r)= \rho_{InSe\_X}(r)-\rho_{InSe}(r)-\rho_X(r)$, where X represents graphene or BN, and $\rho_{InSe\_X}(r)$, $\rho_{InSe}(r)$ and $\rho_X(r)$ are the charge density of the hybrid, InSe, and graphene/BN, respectively. Figures 1d and e show the isosurface plots of the DCD for InSe/graphene and InSe/BN heterostructures, where blue/green denotes depletion/accumulation of electrons. Here, we found a clear flow of electrons transferred from graphene to InSe surface, the total amount of the transferred charge is 0.11 $e$. In contrast, there is an opposite trend of the charge transfer for InSe/BN, where the electron is transferred from InSe to BN (the total amount of the transferred charge is around 0.07 e).

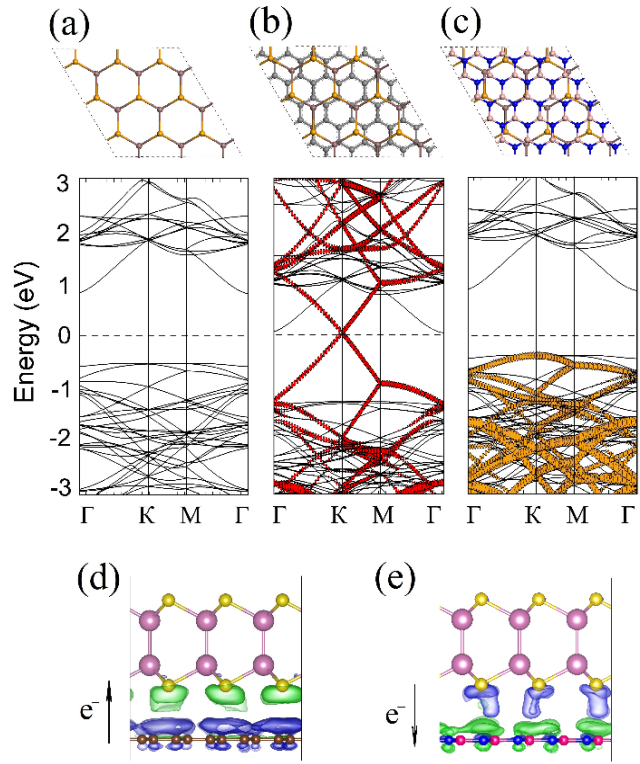

**Figure 1.** Upper panel: the lattice and band structures of (a) InSe, (b) InSe-graphene and (c) InSe-BN. The bands coloured in black, red and orange show InSe, graphene and BN, respectively. The black dashed line shows the Fermi level. Bottom panel: the side view of the 0.12 Å$^{-3}$ DCD isosurface for (d) InSe-graphene and (e) InSe-BN heterostructures. Blue/green denotes depletion/accumulation of electrons, respectively. The spheres coloured in yellow, violet, brown, pink and blue show selenium, indium, carbon, boron and nitrogen atoms. The black arrows show the charge transfer direction.

Our results predict that graphene and BN play an opposite role in doping InSe, which is dramatically different from the phosphorene-graphene/BN bilayer, where graphene and BN play the same role as a weak donor.[24] The charge transfer should be attributed to the dipole-induced charge redistribution at the interface due to the B-N polar bond. The lone-pair electron densities in the Se atom are distorted toward pairing with the N atom, which has a higher electronegativity than the Se atom. Moreover, both B atoms in a BN sheet and Se atoms in a InSe sheet possess Lewis acidic characteristics, thus favouring the interaction between B atoms and Se atoms.[2] It should be noted that the band structure of the InSe/BN bilayer system does not show a typical *p*-type doping behavior of InSe. The underlying reason for that may be a very weak charge transfer from InSe to BN and a weak charge screening effect in this hole-doped 2D system.

We performed additional calculations to examine the stacking effect between the BN/graphene layer with the InSe layer. In the calculations, a relative lateral shift of (0.25, 0.25) is made between the BN/graphene layer and the InSe layer. Our calculations show that there is almost no change in the band structure and the charge transfer, indicating that this effect is minor.

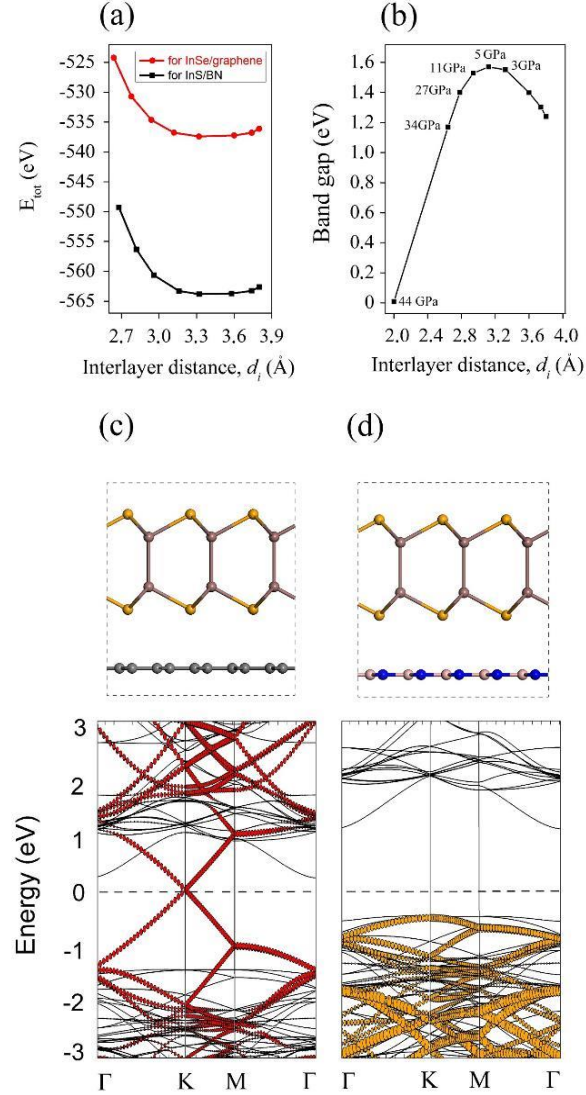

**Figure 2**. (a) The total energy of the system as a function of the interlayer distance $d_i$ in InSe/graphene (red line) and InSe/BN (black line). (b) Modulation of the band gap of the InSe/BN system as a function of the interlayer distance $d_i$. The atomic structure (upper panel) and the band structure (bottom) of (c) InSe/graphene at $d_i = 3.2$ Å and (d) InSe/BN at the $d_i = 3.2$ Å. The bands coloured in black, red and orange represent InSe, graphene and BN, respectively. The black dashed line shows the Fermi level.

In addition to the InSe-graphene/BN bilayer system, we also considered the periodic InSe-graphene/BN superlattice with the sequential stacking of InSe and graphene/BN along the normal direction. As there exist the lone-pair electronic states in the top Se atoms,[20] we are interested in modulating the electronic structure of the hybrids, in particular, to see any indirect-direct transition of InSe, by changing the van der Waals gap. The modified interlayer distance $d_i$ should significantly affect the hybridization of the Se states and the graphene and BN states.

The total energy $E_{tot}$ of InSe/graphene (red line) and that of InSe/BN (black line) systems as a function of $d_i$ are calculated and shown in Figure 2a. The calculated equilibrium distance corresponding to the lowest $E_{tot}$ was found to be around 3.32 Å for both InSe/graphene and InSe/BN structures with the variation of $E_{tot}$ with $d_i$ in a range from 3.2 to 3.5 Å. In Figure 2b, the external pressure corresponding to each $d_i$ below the equilibrium lattice for InSe/BN system is listed. It was found that the InSe/graphene heterostructure remains zero-band gap ($E_g$) for all the considered values of $d_i$ due to the presence of the Dirac cone in graphene. For InSe/BN system, the modification of $d_i$ leads to a nonlinear change of $E_g$. Figure 2b shows that with the decrease of $d_i$, $E_g$ slightly increases from 1.24 eV ($d_i = 3.20$ Å) to the maximum value of 1.57 eV ($d_i = 3.12$ Å), and then monotonically decreases to 0 eV ($d_i = 2.0$ Å). The presence of a concave shape and a strong nonlinearity of the $d_i$ - $E_g$ curve around the equilibrium suggests a strong electron-lattice coupling in the hybrid system. The band structures of InSe/graphene and InSe/BN bulk systems at $d_i = 3.2$ Å are shown in Figures 2d and c, respectively.

Comparing the band structures of bilayer InSe/graphene (Figure 1b), we found that the Dirac cone formed by graphene is also preserved but less shifted to the conduction band edges of InSe (Figure 2c, lower panel). This adjusted band alignment may significantly modify the carrier recombination behaviour at the interface. The optical, electronic and thermal properties under pressure will be dramatically different from those in the freestanding case. For both pressured InSe/graphene and InSe/BN systems, there is no indirect-direct band gap transition in InSe.

**Chemical functionalization of monolayer InSe and InSe/graphene and InSe/BN heterostructures.** We next discuss molecular adsorption of bare InSe and graphene/BN encapsulated InSe. First, we investigated the doping through adsorbing a single K, Al and Mg atom on InSe. The motivation is to induce a *n*-type conduction in InSe as the considered atoms have a small electronegativity and a strong ability to donate electrons. We found that the strongest adsorption site for all considered atoms is above the centre of the hollow hexagon. The electronic properties and the charge transfer are strongly dependent on the distance (*d*) of the dopants to the InSe sheet. Figures 3, 4 and 5 show the band structure and the amount of transferred charge of the K, Al and Mg atoms, respectively, from the surface as a function of the several different separations. For each adsorption case, the *z*-coordinates of the dopants and nearby Se atoms were fixed during calculations. To facilitate the identification of the dopants-induced effect, the band structure of the pristine monolayer InSe was also presented (Figures 3a, 4a and 5a). To clearly show the effect on the Fermi level, the band structures in Figures 3-5 were all adjusted to align the valence band maximum (VBM) of the host InSe. After K adsorption, the bands of InSe remain almost unchanged except a rigid upward shift of the Fermi level (red dashed line), for $d$ = 3.0 (Figure 3b), 3.6 (Figure 3c), and 4.2 Å (Figure 3d), the Fermi level moves into the conduction band.

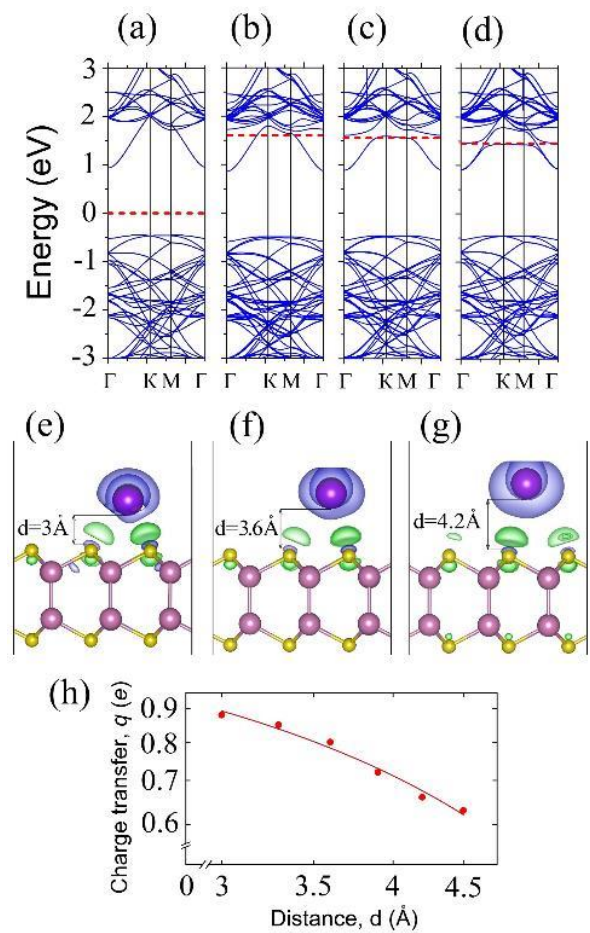

**Figure 3.** The band structures for (a) monolayer InSe structure and InSe with a K atom adsorbed at the distances of (b) 3.0 Å, (c) 3.6 Å and (d) 4.2 Å above the surface. The bands coloured in blue represent InSe. The red dashed line shows the Fermi level. The side view of the isosurface plots of the DCD (the blue/green denotes depletion/accumulation of electrons) for InSe with the K atom adsorbed at the distances of (e) 3.0 Å, (f) 3.6 Å and (g) 4.2 Å. The spheres in yellow, violet and purple show Se, In and K atoms, respectively. (h) The variation of the charge transferred from the K atom to the InSe surface with an increasing distance between the adsorbate and the surface.

Below, we performed quantitative evaluation of the amount of the charge transfer between the K atom and the InSe surface by using DCD $\Delta\rho(r)$, defined as the difference between the total charge density of the K-adsorbed InSe system and the sum of the charge densities of the isolated K atom and the InSe surface. The isosurfaces of the differential charge density for the K atom adsorbed on the InSe surface at $d = 3.0$, 3.6 and 4.2 Å are depicted in Figures 3e-g, respectively. As expected from the band structure analysis, a strong depletion of electrons in the K atom and an accumulation of electrons on the InSe surface was found. In addition, Bader analysis also shows a strong charge transfer from the K atom to the InSe surface with the charge transfer almost linearly increasing with the decrease of $d$ (Figure 3h). The maximum amount of the transferred charge is about 0.88 $e$ (when $d = 3.0$ Å), which is mainly from the $4s^1$ orbital of the K atom.

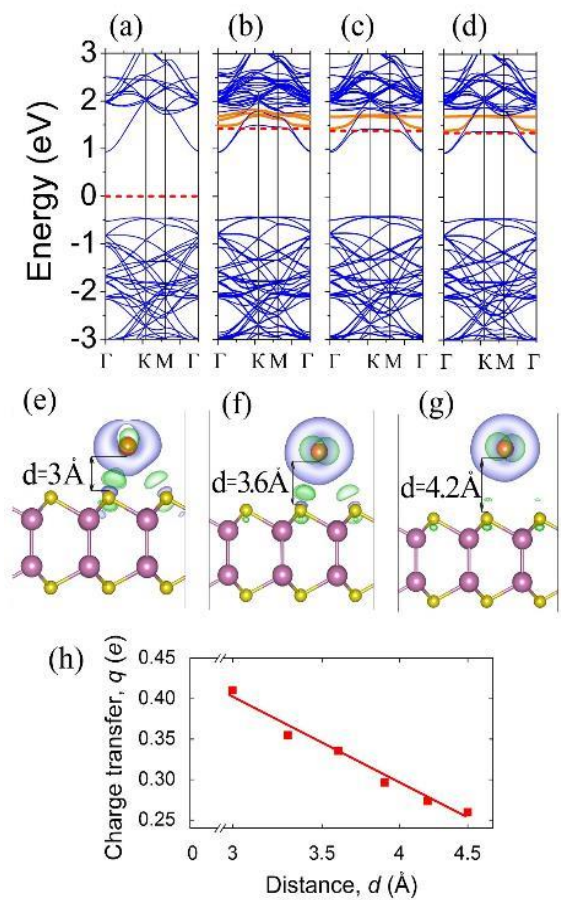

**Figure 4.** The band structures for (a) monolayer InSe structure and InSe with an Al atom adsorbed at the distances of (b) 3.0 Å, (c) 3.6 Å and (d) 4.2 Å above the surface. The bands coloured in blue and orange represent InSe and Al, respectively. The red dashed line shows the Fermi level. The side view of the isosurface plots of the DCD (the blue/green denotes depletion/accumulation of electrons) for InSe with Al atom adsorbed at the distances of (e) 3.0 Å, (f) 3.6 Å and (g) 4.2 Å. The spheres in yellow, violet and orange show Se, In and Al atoms, respectively. (h) The variation of the charge transferred from the Al atom to the InSe surface with increasing the distance between the adsorbate and the surface.

For the case of Al adsorption, we also found that the bands of InSe remain almost unchanged exept a rigid upward shift of the Fermi level (red dashed line) at $d$ = 3.0 (Figure 4b), 3.6 (Figure 4c) and 4.2 Å (Figure 4d), and the Fermi level moves into the conduction band. The isosurfaces of the differential charge density for the Al atom adsorbed on the InSe surface at $d$ = 3.0, 3.6 and 4.2 Å are depicted in Figures 4e, f and g, respectively. Similar to the case of K adsorption, a strong depletion of electrons in Al atom and an accumulation of electrons on InSe surface were found. The Bader analysis predicted a strong electron transfer from the Al atom to the InSe surface with the charge transfer almost linearly increasing with the decrease of $d$ (Figure 4h). The maximum amount of the transferred charge is about 0.41 $e$ (when $d$ =3.0 Å).

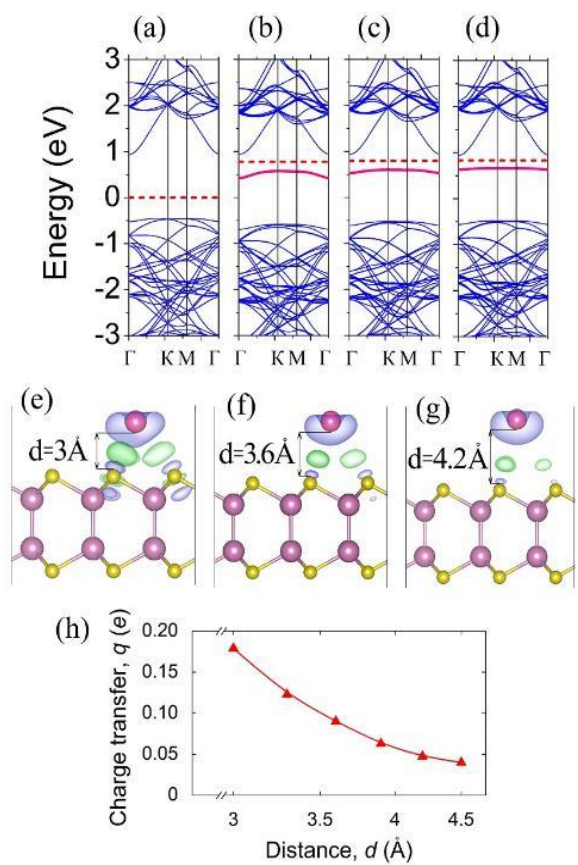

**Figure 5.** The band structures for (a) monolayer InSe structure and InSe with a Mg atom adsorbed at the distances of (b) 3.0 Å, (c) 3.6 Å and (d) 4.2 Å above the surface. The bands coloured in blue and pink represent InSe and Mg, respectively. The red dashed line shows the Fermi level. The side view of the isosurface plots of the DCD (the blue/green denotes depletion/accumulation of electrons) for InSe with the Mg atom adsorbed at the distances of (e) 3.0 Å, (f) 3.6 Å and (g) 4.2 Å. The spheres in yellow, violet and pink show Se, In and Mg atoms, respectively. (h) The variation of the charge transferred from the Mg atom to the InSe surface with an increasing distance between the adsorbate and the surface.

Compared to the effects of the K and Al adsorptions, the Mg adsorption on InSe exerts a different effect on its band structure. Despite the fact that the bands of InSe also remain almost unchanged, the Fermi level (red dashed line) only slightly shifts upward at $d = 3.0$ (Figure 5b), 3.6 (Figure 5c), and 4.2 Å (Figure 5d), and does not cross the conduction band, suggesting that Mg is a weak donor. An additional Mg-induced state appears around 0.7 eV below the conduction band. In addition, our calculations showed that the band gap of Mg-doped InSe is larger than that of bare InSe. More specifically, the band gap for perfect InSe is ~1.35 eV, while for that of InSe with the Mg atom adsorbed at $d$ =3.0, 3.6 and 4.2 Å is ~1.52, 1.46 and 1.42 eV, respectively. It should be noted that in comparing the band gap of perfect InSe with those of Mg-doped InSe, the band gap of Mg-doped InSe was obtained by measuring the VBM and CBM related to the host InSe with the Mg states being excluded.

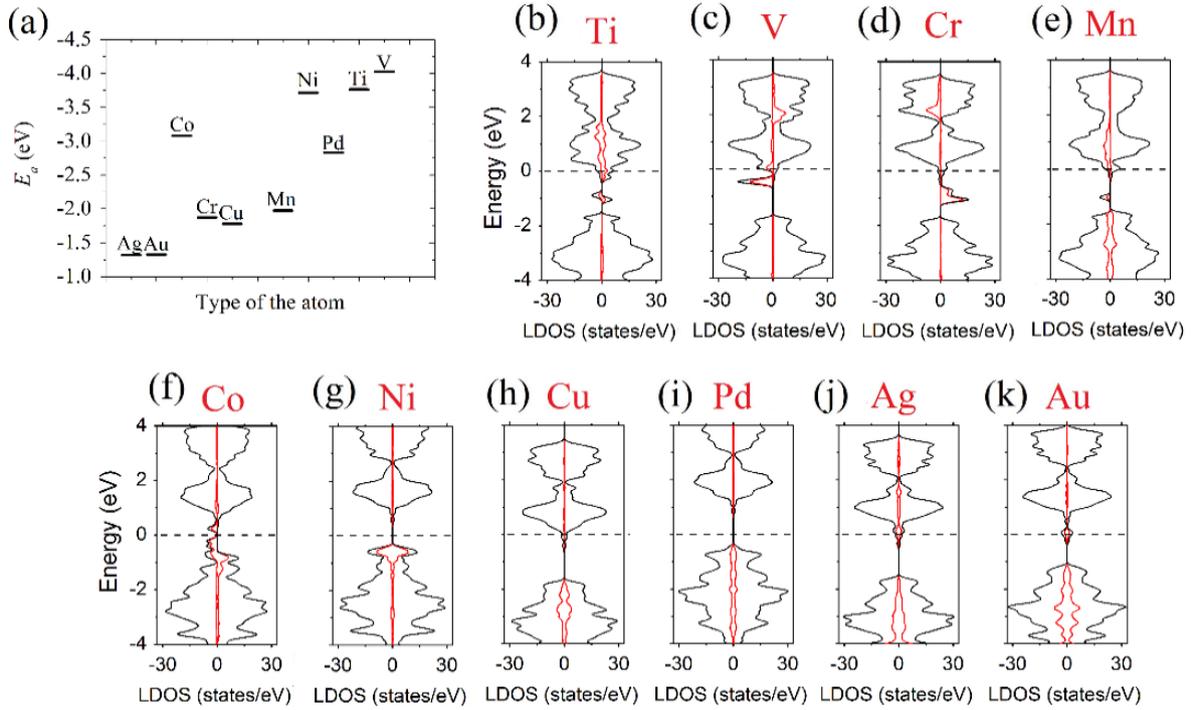

**Figure 6.** (a) Adsorption energies for Ti, V, Cr, Mn, Co, Ni, Cu, Pd, Ag and Au atoms adsorbed on InSe. The local density of states for (b) Ti, (c) V, (d) Cr, (e) Mn, (f) Co, (g) Ni, (h) Cu, (i) Pd, (j) Ag and (k) Au atoms adsorbed on InSe. The LDOS of the adatoms are shown by the red line, the black line represents the total DOS, and the dashed line shows the Fermi level.

The isosurfaces of the DCD for the Mg atom adsorbed on the InSe surface at $d = 3.0$, 3.6 and 4.2 Å are depicted in Figures 5e, f, and g, respectively. A clear depletion of electrons in the Mg atom and an accumulation of electrons on the InSe surface were found. The Bader analysis predicted a charge transfer from the Mg atom to the InSe surface with the value almost quadratically increasing with the decrease of $d$ (Figure 4h). The maximum amount of the transferred charge is about 0.18 $e$ (when $d$ =3.0 Å), which is mainly due to electron redistribution between the Se 4p and the Mg 3s orbitals. A similar charge transfer mechanism for the interaction of Mg with the $Mo_6S_8$ cluster has recently been observed.[40] Such charge transfer from K, Al and Mg dopants may induce an electric field across the dopant/InSe interface. The above finding suggests that the K and Al adsorptions are able to induce a strong *n*-type conduction or even metallicity in InSe.

We also examined the adsorption of most common 3d, 4d and 5d transition metal atoms on InSe. The calculation results showed that the transition metal atoms induce significant lattice distortions of InSe due to the relatively strong interaction. Figure 6a shows the calculated $E_a$ for Ti, V, Cr, Mn, Co, Ni, Cu, Pd, Ag and Au as adatoms on InSe. Figures 6b–k show the spin-polarized LDOS of the adatom–InSe system with Ti, V, Cr, Mn, Co, Ni, Cu, Pd, Ag and Au as adatoms, respectively. The LDOS of Ti-adsorbed InSe (Figure 6b) reflects that Ti is an effective donor with the Fermi level located at the conduction band. There exists a strong hybridization between the Ti and InSe, which is reflected from the highly broadened states. For V-adsorbed system (Figure 6c), the majority spin states cross the Fermi level, while the minority spin states are significantly shifted upward into the conduction band. Therefore, the V-adsorbed InSe has a 100% spin polarization, allowing for spintronic applications. Concerning the Cr-adsorbed InSe (Figure 6d), there are two minority spin peaks below the Fermi level at−1.05 and −0.70 eV, respectively, which are mainly contributed by the 4s states of Cr adatom. A similar magnetic splitting was found in the Mn-doped InSe (Figure 6e). Interestingly, for the Co-doped InSe (Figure 6f), despite the fact that there are several minority spin peaks above and below the Fermi level at 0.25, −0.20, −0.40 and −0.65 eV, the minority spin states have a narrow band gap around the Fermi level, while the majority spin states have a wide band gap around the Fermi level. Figure 6g shows that the Ni-doped InSe system is non-spin-polarized and has a wide band gap around the Fermi level because both the majority and minority spin states are symmetric and situated in the vicinity of the valence band maximum of the host InSe. In the case of the Cu-adsorbed InSe (Figure 6h), the Fermi level shifts into the conduction bands, which causes similar *n*-doping as K. For the Pd case, the Fermi level is located within the gap of InSe, suggesting a weak charge transfer and doping ability. From Figures 6j and k, it can be seen that both Ag and Au are donors since the Fermi level shifts upwardly into the conduction band of InSe.

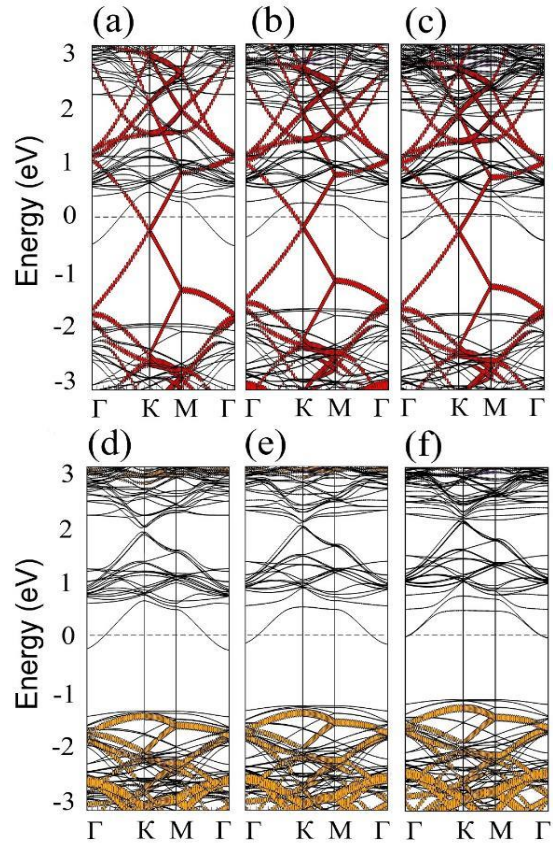

**Figure 7.** The band structures for (a)-(c) InSe/graphene and (d)-(f) InSe/BN heterostructures with the K atom adsorbed at the distances of 3.0, 3.6 and 4.2 Å above the surface, respectively. The bands coloured in black, red, yellow and violet represent InSe, graphene, BN and K atom, respectively. The black dashed line shows the Fermi level.

In addition, we analyzed the adsorption of the K atom on InSe/graphene and InSe/BN heterostructures. Figures 7a-c and Figures 7d-f, respectively, show the band structures of InSe/graphene and InSe/BN heterostructures with the K atom adsorbing at three different heights of $d$ = 3.0, 3.6 and 4.2 Å above the surface (the atomic configurations are presented in Figure 8). As found for monolayer InSe, here we also observed the upward shift of the Fermi level in K-doped InSe/graphene and InSe/BN heterostructures with the decrease of $d$.

The isosurface plots of DCD for K-doped InSe/graphene (Figures 8a-c) and InSe/BN (Figures 8d-f) heterostructures show a strong depletion of electrons in the K atom and an accumulation of electrons on the InSe surface for K at $d$ = 3.0, 3.6 and 4.2 Å. As in the case of monolayer InSe, a linear increase of charge transfer from the K atom to InSe surface with the decrease of $d$ was revealed for both heterostructures (see Figure 8g). The maximum charge transfer from the K atom to both heterostructures reaches up to 0.89 $e$, which is almost the same as that for the monolayer case. This may be due to a strong acceptor ability of the InSe surface, which is able to accumulate charge from the K atom without transferring it to the substrate.

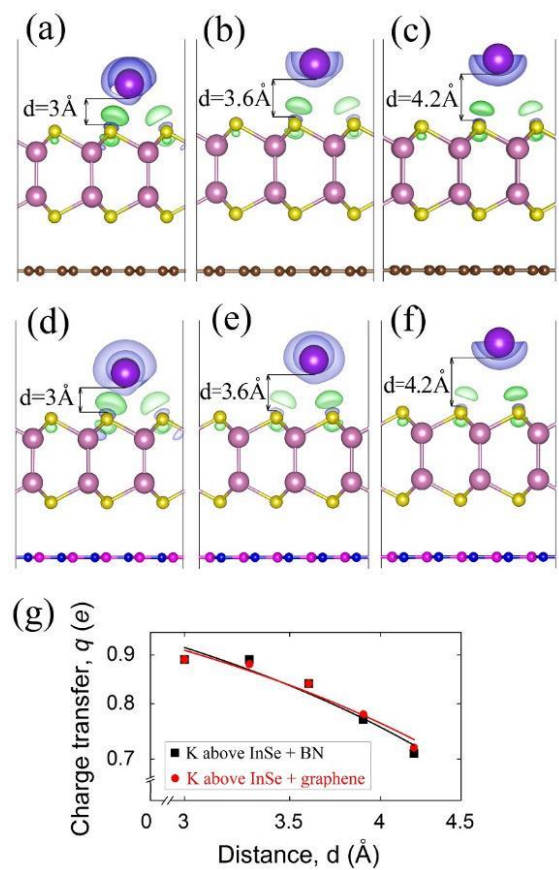

**Figure 8.** The side view of the isosurface plots of the DCD (the blue/green denotes depletion/accumulation of electrons) for (a)-(c) InSe/graphene and (d)-(f) InSe/BN heterostructures with the K atom adsorbed at the distances of 3.0, 3.6 and 4.2 Å above the surface, respectively. The spheres coloured in yellow, violet, brown, pink, blue and purple show selenium, indium, carbon, boron, nitrogen and potassium atoms. (g) The variation of the charge transferred from K atom to InSe/graphene (red line) and InSe/BN (black line) by changing the distance between the atom and the surface.

Another potential approach for chemical functionalization of InSe is through molecular adsorption or atomic functional groups above the surface.[41] Herein, to investigate the effect of typical charge-transfer organic molecules on the electronic properties and the conducting polarity of monolayer InSe, two typical electron-withdrawing molecules TCNE and TCNQ were chosen. The most stable adsorption positions for the TCNE and TCNQ molecules are shown in Figures 9a and b (right panel), respectively, where TCNE (TCNQ) molecule is placed parallel to the InSe surface at a vertical distance of ~3.33 Å (2.91 Å), with the adsorption energy of −0.62 eV (−1.06 eV). The Bader analysis showed the charge transfer of ~0.1 $e$ from the InSe surface to both TCNE and TCNQ molecules. This is comparable with the charge transfer of ~0.1 $e$ from graphene,[32] but about 4 times smaller than the charge transfer of ~0.4 $e$ from phosphorene[33] to these molecules.

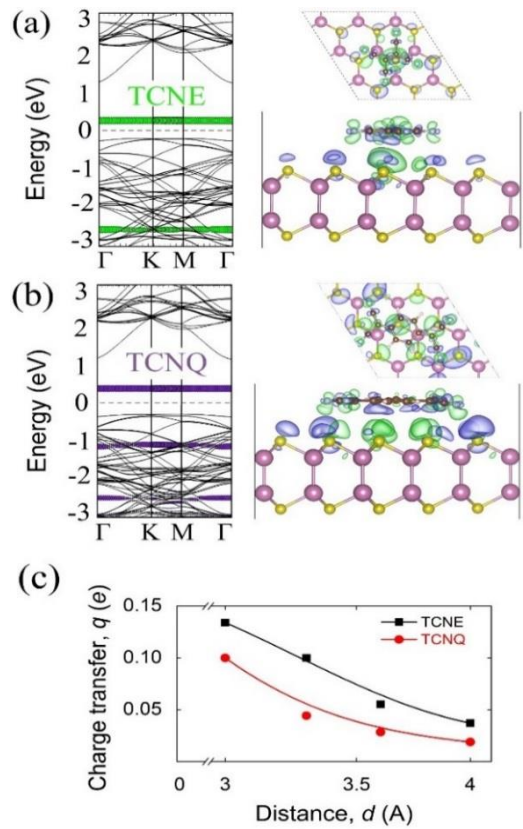

**Figure 9.** The band structures (left panel) and top and side views of the isosurface plots of the DCD (right panel) for monolayer InSe adsorbed with (a) TCNE and (b) TCNQ molecules. For the left panel, bands coloured in black, green and violet represent InSe, TCNE, and TCHQ respectively. The black dashed line shows the Fermi level. For the right panel, the blue/green denotes depletion/accumulation of electrons. Spheres in yellow, violet, brown, blue and pink show selenium, indium, carbon, hydrogen, and nitrogen atoms, respectively. (c) The variation of the charge transferred from the InSe surface to TCNE (black line) and TCNQ (red line) molecules with an increasing distance between molecules and the surface.

Figures 9a and b (left panel) show the band structures for TCNE- and TCNQ-adsorbed InSe surfaces. In both cases, a flat level originating from the lowest unoccupied molecular orbital (LUMO) of the adsorbed molecules appears within the gap and is close to the VBM. This shallow unoccupied level allows the transfer of electrons from InSe. Furthermore, the charge transfer is strongly dependent on $d$ between molecules and the InSe surface. The Bader analysis predicted a decrease in the charge transfer from the InSe surface to both molecules with an increase of the adsorption distance (Figure 9c).

**Conclusions**

By using first-principles calculations, we investigated the modulation of monolayer InSe electronic properties by encapsulation, surface functionalization, and molecular adsorption. In particular, we showed an opposite charge donating role of graphene (donor) and BN (acceptor) in InSe, which is dramatically different from phosphorene, where both graphene and BN play the same role (donor). A decrease in interlayer spacing between InSe-graphene/BN is able to dramatically change the bands alignment and tune the band gap.

We also studied the effects of surface doping with K, Al, Mg and typical transition metal atoms, as well as TCNE and TCNQ organic molecules on the band structure and charge transferability of monolayer InSe. For the K-doped monolayer InSe, we predict a semiconductor to metal transition and a linearly increased charge transfer from the dopant to the InSe surface with a decreasing distance between them. For the strong atomic donors like K and Al, the charge transfer across the dopant-InSe interface is accompanied with a strong redistribution of the electric potential or field at the interface, which may alter the photon kinetics and electron-hole recombination efficiency of InSe. In fact, the effect of the Al-induced field redistribution on light-emission/adsorption has been observed in phosphorene.[42] In addition, adsorption of InSe by the transition metal atoms induces significant lattice distortions, which results in changes of the electronic properties of InSe. Interestingly, spin polarization occurs upon Ti, V, Cr, Mn and Co adsorptions. Our present study not only reveals insights into the modulation of electronic properties of free-standing and graphene/BN substrate-supported InSe, but also renders new ways to control its electronic structure and carrier density, which may pay the way for the practical applications of InSe in nanoelectronics and gas sensors.

**Acknowledgements**

The authors acknowledge the financial support from the Agency for Science, Technology and Research (A*STAR), Singapore, and the use of computing resources at the National Supercomputing Centre, Singapore. This work was supported in part by a grant from the Science and Engineering Research Council (152-70-00017) and the Ministry of Education, Singapore (Academic Research Fund TIER 1—RG128/14). S. V. Dmitriev acknowledges financial support from the Russian Science Foundation Grant N 14-13-00982.